\documentclass[journal=jacsat,manuscript=article]{achemso}

\usepackage[version=3]{mhchem} 
\usepackage{xcolor}
\usepackage{adjustbox}
\usepackage{graphics}
\usepackage{adjustbox}

\title{Data-driven Design of High Pressure Hydride Superconductors using DFT and Deep Learning}%

\author{Daniel Wines}%
 \email{daniel.wines@nist.gov}
 \affiliation{%
 Material Measurement Laboratory, National Institute of Standards and Technology,
Gaithersburg, MD 20899, USA 
}%

\author{Kamal Choudhary}
 \affiliation{%
 Material Measurement Laboratory, National Institute of Standards and Technology,
Gaithersburg, MD 20899, USA 
}%

\begin{document}

\begin{abstract}

The observation of superconductivity in hydride-based materials under ultrahigh pressures (for example, H$_3$S and LaH$_{10}$) has fueled the interest in a more data-driven approach to discovering new high-pressure hydride superconductors. In this work, we performed density functional theory (DFT) calculations to predict the critical temperature ($T_c$) of over 900 hydride materials under a pressure range of (0 to 500) GPa, where we found 122 dynamically stable structures with a $T_c$ above MgB$_2$ (39 K). To accelerate screening, we trained a graph neural network (GNN) model to predict $T_c$ and demonstrated that a universal machine learned force-field can be used to relax hydride structures under arbitrary pressures, with significantly reduced cost. By combining DFT and GNNs, we can establish a more complete map of hydrides under pressure. 

\end{abstract}


\maketitle
\section{Introduction}
The discovery of superconductors with a high transition temperature ($T_c$) has been a long-standing goal in the condensed matter physics community. The highest measured $T_c$ for a Bardeen-Cooper-Schrieffer (BCS)\cite{PhysRev.106.162,PhysRev.108.1175} conventional superconductor (mediated by electron-phonon interactions) in ambient conditions has been measured to be 39 K for MgB$_2$ \cite{mgb2}. These moderately low $T_c$ values (significantly less than liquid nitrogen) of traditional BCS superconductors can be overcome by applying ultra-high pressures to hydrogen-rich compounds \cite{hydride-review,hydridereview2}. Metallic hydrogen and hydrogen-rich compounds are ideal for high temperature superconductivity because hydrogen atoms provide strong electron-phonon coupling (EPC) and high frequency phonon modes \cite{PhysRevLett.21.1748,PhysRevLett.92.187002}. Recent experimental observation (within the past decade) of conventional superconductivity in hydride-based materials under high pressure has reinvigorated the interest in searching for new hydride superconductors and minimizing the required pressure to sustain superconductivity at a high temperature \cite{hydride-review,hydridereview2}. 

This renowned interest in high pressure hydrides was reignited in 2015 with the experimental measurement of superconductivity at 203 K for H$_3$S under pressure \cite{h3s}. In addition to H$_3$S, a $T_c$ above 250 K has been experimentally observed for LaH$_{10}$ above 170 GPa \cite{lah10,https://doi.org/10.1002/anie.201709970,PhysRevLett.122.027001}, with several experimental and computational studies confirming these results \cite{lah10,doi:10.1073/pnas.1704505114,https://doi.org/10.1002/anie.201709970,PhysRevLett.122.027001,PhysRevB.98.100102,PhysRevB.101.024508}. A $T_c$ of 243 K has been observed for YH$_9$ at 201 GPa \cite{yh9} and superconductivity has also been measured for ThH$_{10}$ \cite{SEMENOK202036}, YH$_4$ \cite{PhysRevB.104.174509}, LuH$_3$ \cite{luh3}, zirconium polyhydrides \cite{ZHANG2022907}, and tin hydrides \cite{HONG2022100596} under pressure. More recently, calcium superhydrides such as CaH$_6$ have been measured to have a $T_c$ of up to 210 K at (160 to 190) GPa \cite{cah6}. It has also been demonstrated that doping can play a role in enhancing $T_c$ for hydride materials under pressure \cite{denchfield2023novel,PhysRevB.104.214505,GE2020100330,FAN2016105}. Recent claims of room temperature superconductivity in Ca-H-S (under 190 GPa) \cite{dias-Ca-H-S} and Lu-H (under 10 GPa) \cite{dias-Lu-H} systems have been met with scrutiny from the scientific community, with Ref. \cite{dias-Ca-H-S} and Ref. \cite{dias-Lu-H} being retracted. Despite the lack of reproducibility in these recent claims, the search for new hydride superconductors with a high $T_c$ continues on the experimental and computational front.     

In contrast to unconventional superconductors, where there is not a well-defined theoretical framework to describe superconductivity, BCS theory can be utilized as a tool to screen potential superconducting materials in conjunction with electronic structure methods such as density functional theory (DFT). With the advancements in computing power, DFT databases, and robust and efficient computational workflows, high-throughput calculations (on the order of hundreds to thousands) of superconducting properties are now in the realm of possibility \cite{doi:10.1063/1.4812323,oqmd,oqmd-2,Haastrup_2018,Gjerding_2021,choudhary2020joint,ONG2013314,10.1063/5.0159299}. In addition, various machine learning and data-driven approaches have been used to aid in the discovery of new and novel superconductors \cite{bulksc,2dsc,ml-sc1,ml-sc2,ml-sc3,C8ME00012C,ROTER20201353689,ml-sc4,SEEGMILLER2023112358,inverse-sc,PhysRevB.104.054501,sommer20223dsc,cerqueira2023sampling}. In our previous works, we expanded the existing JARVIS (Joint Automated Repository for Various Integrated Simulations) \cite{10.1063/5.0159299,choudhary2020joint} DFT database to include calculations for over 1000 bulk \cite{bulksc} and 150 two-dimensional (2D) superconductors \cite{2dsc} (all at zero pressure). We also developed a deep-learning tool (trained on our DFT dataset) to predict the electron-phonon coupling (EPC) parameters and T$_c$ of any material \cite{bulksc} at a significantly reduced cost. We went on to utilize this deep-learning tool for forward and inverse design of new superconductors \cite{bulksc,inverse-sc}. 

In addition to our own work, there has been a recent high-throughput DFT and machine learning study which performed EPC calculations for over 7000 superconducting candidates \cite{cerqueira2023sampling} (all at zero pressure), which makes it one of the largest high-throughput DFT study of conventional superconductors to date. In contrast to the DFT and machine learning works that span several material classes, there have been few high-throughput computational efforts that have focused on high pressure hydrides \cite{PhysRevB.104.054501}. Recently, the Superhydra database was created by Saha et al. \cite{PhysRevMaterials.7.054806} to map superconductivity in high pressure hydrides, but the DFT calculations have so far been limited to binary compounds at a single pressure value of 200 GPa. In addition, Belli et al. \cite{bondingnetwork} analyzed the electronic and structural properties of more than 100 hydride superconductors calculates with DFT (at several pressures ranging from 0 GPa to 500 GPa) and found a strong correlation between $T_c$ and electronic bonding network, establishing a clear path to engineer the $T_c$ \cite{bondingnetwork}. In a seminal work in 2021, Shipley et al. \cite{PhysRevB.104.054501} surveyed the landscape of binary hydrides up until 500 GPa for the entire periodic table and performed high-throughput DFT calculations for the best candidates at pressure values of 10, 100, 200, 300 and 500 GPa. From these results, they found 36 dynamically stable high pressure hydride superconductors with a $T_c$ above 100 K, with 18 of these materials having a $T_c$ over 200 K \cite{PhysRevB.104.054501}.

Building on our previous efforts within JARVIS and inspired by the recent efforts of other researchers focused on hydrogen-based superconductors, we established a high-throughput workflow to screen potential hydride superconductors. Specifically, we extended our search beyond binary compounds. We ran DFT calculations for over 900 hydrogen-based materials. These calculations involved relaxing the hydride structures under various amounts of pressure and then performing EPC calculations to determine superconducting properties. We went on to further analyze dynamically stable structures (containing all positive phonon frequencies in the phonon density of states) with the highest $T_c$ and the structures that were not previously known to be superconducting. After running these DFT calculations, we trained a deep learning model (using the relaxed high-pressure structures and computed $T_c$ values) to predict $T_c$ given a crystal structure, specifically tailored to high pressure hydrides. Such deep learning models have shown remarkable potential for superconductor design\cite{choudhary2022recent,bulksc,choudhary2024atomgpt,cerqueira2023sampling,burdine2023discovery,inverse-sc}. After testing and benchmarking our newly trained deep learning property prediction model, we used a unified machine learning-based force-field to relax the hydride structures under pressure, further reducing the number of DFT calculations needed to obtain a complete map of superconducting properties under pressure. 

\section{Methodology}
In order to compute the EPC and $T_c$ of hydride materials, we performed non-spin polarized density functional perturbation theory (DFPT) \cite{baroni1987green,gonze1995perturbation} with the Gaussian broadening (interpolation) method \cite{wierzbowska2005origins}. Similarly to our previous workflows in Ref. \cite{bulksc}, Ref. \cite{2dsc}, and Ref. \cite{inverse-sc}, we used the Quantum Espresso \cite{giannozzi2020quantum} software package, Garrity-Bennett-Rabe-Vanderbilt (GBRV) pseudopotentials \cite{garrity2014pseudopotentials} and the Perdew-Burke-Ernzerhof functional revised for solids PBEsol \cite{perdew2008restoring}. For Lu, a Topsakal-Wentzcovitch \cite{TOPSAKAL2014263} pseudopotential was used. Due to the high-throughput nature of this study, which is primarily focused on screening potential high-pressure hydride candidates, we used preconverged k-points from the JARVIS database \cite{choudhary2019convergence}, a 2x2x2 q-point grid and a kinetic energy cutoff of 610 eV (45 Ry). It was demonstrated in Ref. \cite{bulksc} and Ref. \cite{2dsc} that these lower q-point grids can still be effective for material screening purposes. For structures not already in the JARVIS dataset (materials added from the literature), we used the automated k-point convergence scheme in JARVIS \cite{choudhary2019convergence}.

The Eliashberg spectral function (${\alpha}^2 F(\omega)$) \cite{MARSIGLIO2020168102} that determines the EPC can be obtained from:

\begin{equation} 
{\alpha}^2 F(\omega)=\frac{1}{2{\pi}N({\epsilon_F})}\sum_{qj}\frac{\gamma_{qj}}{\omega_{qj}}\delta(\omega-\omega_{qj})w(q)
\end{equation} 
where $N({\epsilon_F})$ is the density of states (DOS) at the Fermi level ${\epsilon_F}$, $\omega_{qj}$ is the mode frequency, $\delta$ is the Dirac-delta function, $w(q)$ is the weight of the $q$ point, and $\gamma_{qj}$ is the phonon mode $j$ linewidth at wave vector $q$:

\begin{equation} 
\gamma_{qj}=2\pi \omega_{qj} \sum_{nm} \int \frac{d^3k}{\Omega_{BZ}}|g_{kn,k+qm}^j|^2 \delta (\epsilon_{kn}-\epsilon_F) \delta(\epsilon_{k+qm}-\epsilon_F)
\end{equation} 
The integral is over the first Brillouin zone, $g_{kn,k+qm}^j$ is the electron-phonon matrix element, and $\epsilon_{kn}$ and $\epsilon_{k+qm}$ are the DFT eigenvalues with wavevector $k$ and $k+q$ within the $n$th and $m$th bands. The relation between $\gamma_{qj}$ and the mode EPC parameter is:

\begin{equation} 
\lambda_{qj}=\frac {\gamma_{qj}}{\pi hN(\epsilon_F)\omega_{qj}^2}
\end{equation} 
The EPC parameter can now be written as:

\begin{equation} 
\lambda=2\int \frac{\alpha^2F(\omega)}{\omega}d\omega=\sum_{qj}\lambda_{qj}w(q)
\label{eq:lambda}
\end{equation} 
where $w(q)$ is the weight of a $q$ point. The McMillan-Allen-Dynes \cite{mcmillan1968transition} formula can be used to approximate the superconducting transition temperature ($T_c$) and is written as:

\begin{equation}
T_c=\frac{\omega_{log}}{1.2}\exp\bigg[-\frac{1.04(1+\lambda)}{\lambda-\mu^*(1+0.62\lambda)}\bigg]\label{eq:mad}
\end{equation}
where
\begin{equation} 
\omega_{log}=\exp\bigg[\frac{\int d\omega \frac{\alpha^2F(\omega)}{\omega}\ln\omega}{\int d\omega \frac{\alpha^2F(\omega)}{\omega}}\bigg]
\end{equation} 
In Eq.~\ref{eq:mad}, the constant $\mu^*$ is the effective Coulomb potential parameter, which we take to be 0.1. To generate our DFT dataset, it took roughly 11,500 core hours. Approximately 7 $\%$ of the computational time was devoted to relaxing the structures under pressure while the remaining 93 $\%$ of the computational time was devoted to DFPT calculations of the EPC.

To predict the superconducting properties of each high pressure hydride material, we used the atomistic line graph neural network (ALIGNN) \cite{choudhary2021atomistic}. Specifically, we trained an ALIGNN model on our new high pressure hydride DFT dataset, and trained an ALIGNN model on a combination of our high pressure hydride DFT dataset and our bulk superconducting DFT dataset from Ref. \cite{bulksc}. It is important to note that both of these ALIGNN models were trained on DFT relaxed structures under pressure. Within ALIGNN, the crystal structure is represented as a graph (elements are nodes, bonds are edges). Each node in the graph has nine input features, which include block, electronegativity, valence electrons, group number, electron affinity, ionization energy, atomic volume, and covalent radius. The bond distances are the edge features and the radial basis function cutoff is 8 $\textrm{\AA}$. In addition, a periodic graph construction with 12-nearest neighbors is used. The line graph is constructed from the atomistic graph, using bond distances as nodes and bond angles as edges. The node and edge features are updated with edge-gated graph convolution using a propagation function. One layer is composed of an edge-gated graph convolution on the bond graph and an edge-gated convolution on the line graph. The line graph convolution produces bond messages that propagate to the atomistic graph, where the atom and bond features are updated further. For superconducting transition temperature, we used 300 epochs for training, a 80:10:10 split and a batch size of 16 (the test set was not used at all during training). The hyperparameters were kept the same as the original ALIGNN paper by Choudhary and DeCost \cite{choudhary2021atomistic}. ALIGNN was designed to be a flexible framework that could be applied to any material class for an incredibly diverse range of material properties. In fact, this formalism has been tested on over 52 properties for materials and molecules that range the entire periodic table, with great success \cite{choudhary2021atomistic}. Extensive benchmarking and testing of model parameters such as node features, optimized hyperparameters, number of epochs and batch size are presented in Ref. \cite{choudhary2021atomistic}. For consistency and due to the fact that we have a similar dataset size \cite{bulksc}, we used the same number of epochs and the same batch size as our previous superconducting work \cite{bulksc}, in addition to the same set of hyperparameters. In addition to the several follow-up studies which use ALIGNN \cite{10.1063/5.0159299} (including our previous superconducting work \cite{bulksc}) which used the same nine node features, we used the same nine node features in the current work. It is entirely possible that adding an additional descriptor to ALIGNN can improve the model to cater to the characteristics of BCS superconductors under pressure, but that is something beyond the scope of this work. ALIGNN is implemented in PyTorch \cite{paszke2019pytorch} and the deep graph library \cite{wang2019deep}. To train our ALIGNN models for our newly generated hydride dataset and a combination of our hydride dataset and previous dataset, it roughly took 4 and 7 GPU hours, respectively.


\section{Results and Discussion}

\subsection{High-throughput DFT Results}

\begin{figure}[h]
    \centering
    \includegraphics[trim={0. 0cm 0 0cm},clip,width=0.7\textwidth]{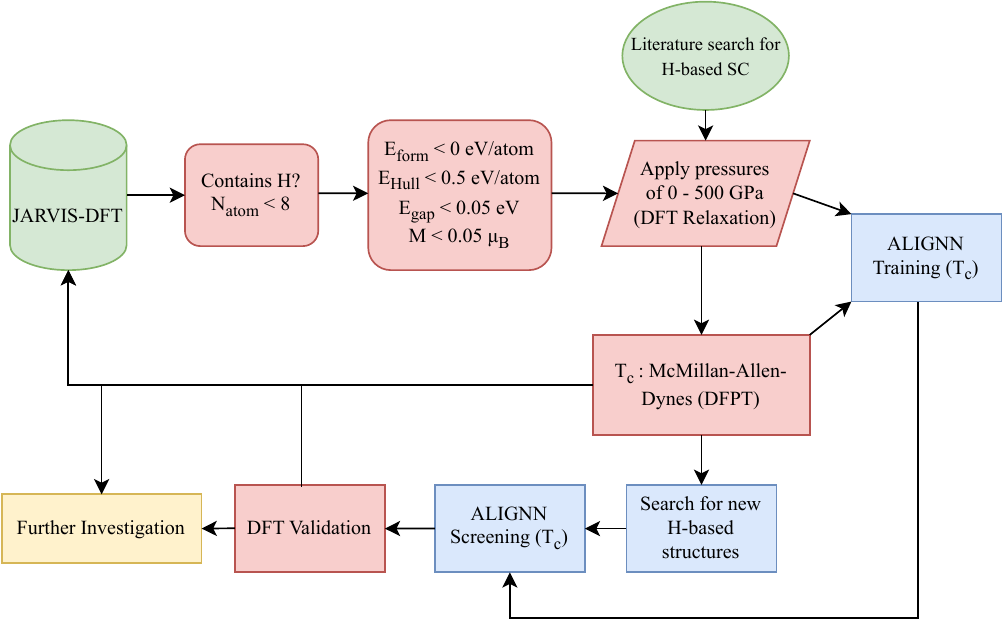}
    \caption{The full workflow for new high pressure hydride superconductors using JARVIS, DFT, and ALIGNN. This workflow involves screening structures from the JARVIS-DFT database (filtering structures that contain H, are thermodynamically stable, nonmagnetic and metallic) and adding structures from literature, then performing DFT calculations at applied pressures of 0 GPa to 500 GPa and computing $T_c$. The ALIGNN model is then trained on DFT data to predict $T_c$ with lower cost, enabling enhanced screening of candidate superconductors. }
    \label{hydride-workflow}
\end{figure}

Fig. \ref{hydride-workflow} depicts our full high-throughput workflow used to investigate high pressure hydride superconductors. The first step involves screening the existing JARVIS-DFT database, which contains over 80,000 materials and millions of computed properties. The first task was to screen materials that contain hydrogen and have a unit cell of 8 or less atoms (structures where the EPC can feasibly be computed, with the exception of certain binary superhydrides such as LaH$_{10}$). After this first round of screening, we screened certain DFT computed properties from JARVIS-DFT such as the formation energy per atom (E$_{\textrm{form}}$), the energy above the convex hull (E$_{\textrm{Hull}}$), the electronic band gap (E$_{\textrm{gap}}$), and the magnetic moment per atom (M). To ensure the superconducting candidate materials were the most thermodynamically stable, metallic and nonmagnetic, the criteria we established (similar to our previous work \cite{bulksc,2dsc,inverse-sc}) was E$_{\textrm{form}}$ $<$ 0 eV/atom, E$_{\textrm{Hull}}$ $<$ 0.5 eV/atom, E$_{\textrm{gap}}$ $<$ 0.05 eV, and M $<$ 0.05 $\mu_{\textrm{B}}$. These properties in the JARVIS-DFT database are calculated with the Opt-b88-vdW \cite{Klimes_2010} functional, which tends to underestimate the band gap. This band gap underestimation is not of concern for screening purposes since we only wish to identify metallic structures. We chose such a generously high E$_{\textrm{Hull}}$ cutoff due to the possibility of applied pressure stabilizing structures (it is possible that a structure at 0 GPa could lie very far above the hull but at higher pressure could lie on the hull). In addition to pulling candidate structures from JARVIS-DFT, we added candidate structures from literature. These included H-based structures from the 3DSC \cite{sommer20223dsc} database. In addition, several binary compounds from the work of Shipley et al. \cite{PhysRevB.104.054501} were included in our calculations. After identifying the potential hydride superconductors, we applied pressures of (0, 100, 200, 300 and 500) GPa to each material and relaxed the structure. To be consistent with our previous work \cite{bulksc}, we performed our new DFT calculations (taking H-based structures from JARVIS, 3DSC \cite{sommer20223dsc} and Shipley et al. \cite{PhysRevB.104.054501} with the same workflow and settings as in Ref. \cite{bulksc} (same PBEsol functional, pseudopotential, etc.). After obtaining the final structure under pressure, we performed DFPT calculations to obtain the EPC. then used the McMillan-Allen-Dynes equation to estimate $T_c$. In addition to obtaining results for over 900 EPC calculations for $T_c$, we used our DFT data to train an ALIGNN model from scratch to predict $T_c$ specifically for hydride structures under pressure. By obtaining a well-trained deep learning model for $T_c$, we can eliminate the need for expensive DFT calculations. Specifically, we can use ALIGNN as a screening tool to identify potential candidates outside of the training set, which we can verify with DFT (see Fig \ref{hydride-workflow}).

\begin{figure}[h]
    \centering
    \includegraphics[trim={0. 0cm 0 0cm},clip,width=1.0\textwidth]{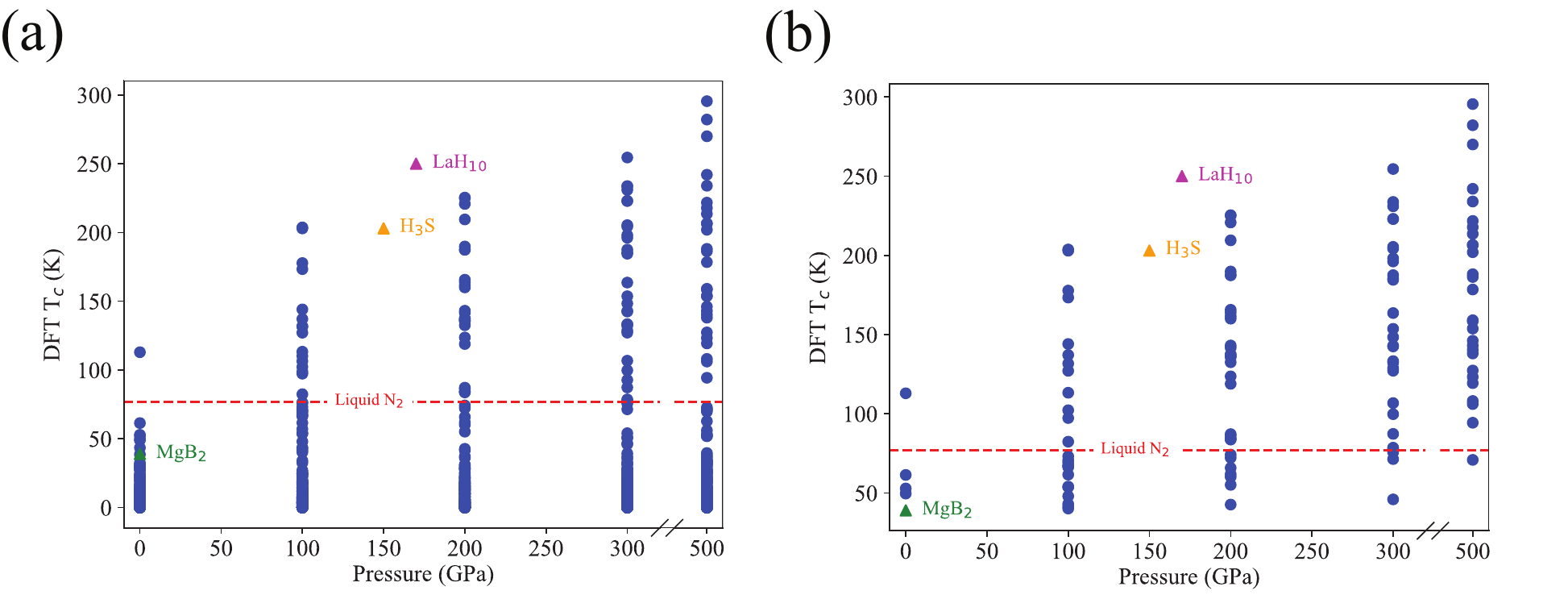}
    \caption{The DFT calculated $T_c$ for the entire dataset at applied pressures of 0 - 500 GPa. For reference, experimental values for MgB$_2$, H$_3$S and LaH$_{10}$ are given. The dotted line represents the boiling point of liquid N$_2$ at ambient pressure. a) Depicts all of the DFT results while b) depicts the dynamically stable structures with a $T_c$ above MgB$_2$ (39 K).  }
    \label{hydride-full}
\end{figure}

Fig. \ref{hydride-full}a) depicts a full summary of the over 900 DFT calculations performed in this study. These calculations were performed for the same set of materials at (0, 100, 200, 300 and 500) GPa (500 GPa was selected as an example of extreme pressure). Fig. \ref{hydride-full}b) depicts a subset of Fig. \ref{hydride-full}a), which includes the dynamically stable structures with a $T_c$ above MgB$_2$ (39 K). Fig. \ref{hydride-full} also depicts reference experimental values for MgB$_2$ (39 K at ambient pressure), H$_3$S (203 K at 150 GPa) and LaH$_{10}$ (250 K at 170 GPa) and the dotted line represents the boiling point of liquid N$_2$ at ambient pressure (77 K). Table \ref{table:1} and Table S1 (supplementary) display the contents of Fig. \ref{hydride-full}b) in tabulated form, with additional details of where the initial structure originated before external pressure was applied in our DFT calculations. It is important to note that a few simulations were performed at intermediate pressures of 50 GPa and 250 GPa. These are included in the tables but omitted from Fig. \ref{hydride-full} for visual purposes. 

At this stage, it is important to discuss the successes and limitations of our theoretical calculations within the context of this work. The goal of this work is to use high-throughput DFT to 1) identify new high pressure hydride superconductors and 2) generate a substantial amount of data to train a reliable deep learning model to predict $T_c$. Due to the high-throughput nature of these calculations and limitations to the theoretical framework, accuracy of the results may be impacted. There are two main sources of variation that can arise when calculating $T_c$. The first type of potential variation stems from the underlying DFT calculations, where calculation parameters such as the size of the q-point grid, choice of exchange-correlation functional, and choice of pseudopotential can impact the accuracy of the result. The second type of variation can occur from the theoretical framework used to estimate the $T_c$. Two types of approaches are normally used to estimate $T_c$ from the underlying DFT results for EPC, either solving the Migdal-Eliashberg equations directly \cite{osti_7354388} or plugging the DFPT computed $\lambda$ and $\omega_{\textrm{log}}$ into the McMillan-Allen-Dynes equation (as we did in this work, Eq.~\ref{eq:mad}). Both of these approaches utilize the empirical constant $\mu^*$, which can introduce variations in the $T_c$ results. In our previous work \cite{bulksc,2dsc}, we found that varying $\mu^*$ from 0.03 to 0.18 can change the $T_c$ by up to 40 $\%$, depending on the material, with the variation being more apparent for structures with a higher $T_c$. At larger EPC strengths, the variation between results obtained from the Migdal-Eliashberg equation and results obtained with the McMillan-Allen-Dynes equation begin to diverge, with the McMillan-Allen-Dynes approach underestimating the $T_c$ \cite{PhysRevB.104.054501,doi:10.1073/pnas.1704505114}. In fact, it was demonstrated that the variations arising from $\mu^*$ combined with the variations arising from the Migdal-Eliashberg vs. McMillan-Allen-Dynes approach can be up to 100 K for materials with large $\lambda$ \cite{PhysRevB.104.054501}. An example of this includes the underestimation of our DFT results vs. experimental values for the $T_c$ of H$_3$S and LaH$_{10}$ (we computed a max $T_c$ of (160 - 187) K at (200 - 250) GPa for LaH$_{10}$ and a $T_c$ of 154 K at 200 GPa for H$_3$S). Nonetheless, important qualitative trends can be identified from these calculations and previously unidentified superconductors can be revealed. This can motivate future, more in-depth studies of selected superconducting materials, with the goal of achieving quantitative experimental accuracy for a smaller, focused subset of materials (rather than screening hundreds, or thousands of candidates with a less computationally costly approach). In order to achieve even higher accuracy and eliminate empirical biases, methods such as Superconducting-DFT and explicit calculations of $\mu^*$ with the random phase approximation can be performed \cite{scdft,PhysRevB.72.024546,PhysRevB.72.024545}.  

\begin{table}[]
\label{table1}
\begin{adjustbox}{width=0.55\textwidth}
\small
\begin{tabular}{l|l|l|l|l}
Structure    & Pressure (GPa) & $T_c$ (K) & E$_{\textrm{form}}$ (eV/atom) & Source of Structures \\
\hline
MgH$_{12}$     & 500 & 296 & -10.9 & Shipley et al. \\
NaH$_6$        & 500 & 282 & -14.4 & Shipley et al. \\
SrH$_{10}$     & 500 & 270 & -13.2 & Shipley et al. \\
MgH$_{12}$     & 300 & 255 & -6.7  & Shipley et al. \\
NaH$_6$        & 500 & 242 & -12.0 & Shipley et al. \\
NaH$_6$        & 300 & 234 & -4.0  & Shipley et al. \\
NaH$_9$        & 500 & 234 & -11.3 & Shipley et al. \\
MgH$_6$        & 300 & 231 & -7.5  & Shipley et al. \\
NaH$_6$        & 200 & 225 & -5.4  & Shipley et al. \\
LiH$_6$        & 300 & 223 & -6.4  & Shipley et al. \\
LiH$_3$        & 500 & 222 & -10.9 & Shipley et al. \\
MgH$_6$        & 200 & 221 & -3.4  & Shipley et al. \\
MgH$_{10}$     & 500 & 218 & -11.1 & Shipley et al. \\
MgH$_6$        & 500 & 214 & -11.8 & Shipley et al. \\
CaH$_6$        & 200 & 210 & -5.9  & Shipley et al. \\
LiH$_6$        & 500 & 207 & -10.5 & Shipley et al. \\
LiH$_2$        & 300 & 205 & -7.2  & Shipley et al. \\
NaH$_6$        & 100 & 204 & -1.9  & Shipley et al. \\
MgH$_{10}$     & 300 & 204 & -6.9  & Shipley et al. \\
LiH$_2$        & 500 & 202 & -11.2 & Shipley et al. \\
CaH$_6$        & 300 & 198 & -8.3  & Shipley et al. \\
YH$_9$         & 300 & 196 & -5.7  & Shipley et al. \\
LiH$_2$        & 200 & 190 & -3.1  & Shipley et al. \\
LiH$_6$        & 200 & 188 & -3.6  & Shipley et al. \\
LiH$_3$        & 300 & 188 & -6.9  & Shipley et al. \\
YH$_9$         & 500 & 188 & -9.5  & Shipley et al. \\
LaH$_{10}$     & 250 & 187 & -4.7  & JVASP-149370   \\
KH$_{10}$      & 500 & 186 & -15.6 & Shipley et al. \\
LaH$_{10}$     & 300 & 185 & -5.7  & JVASP-149370   \\
NaH$_9$        & 300 & 185 & -7.2  & Shipley et al. \\
CaH$_{15}$     & 500 & 179 & -11.3 & Shipley et al. \\
CaH$_6$        & 100 & 178 & -3.5  & Shipley et al. \\
MgH$_{13}$     & 100 & 173 & -1.8  & Shipley et al. \\
H$_3$S         & 200 & 166 & -5.7  & JVASP-79487
\\
MgH$_{13}$     & 200 & 164 & -3.6  & Shipley et al. \\
CaH$_{15}$     & 300 & 164 & -6.8  & Shipley et al. \\
LiH$_3$        & 200 & 162 & -3.4  & Shipley et al. \\
LaH$_{10}$     & 200 & 160 & -3.8  & JVASP-149370   \\
ScH$_{12}$     & 500 & 159 & -11.2 & Shipley et al. \\
Na$_2$H$_{11}$ & 500 & 159 & -12.1 & Shipley et al. \\
H$_3$S         & 250 & 154 & -3.9  & JVASP-79487    \\
KH$_{10}$      & 300 & 154 & -9.3  & Shipley et al. \\
Na$_2$H$_{11}$ & 500 & 154 & -12.1 & Shipley et al. \\
CaH$_{10}$     & 300 & 148 & -7.3  & Shipley et al. \\
ScH$_{14}$     & 500 & 146 & -11.0 & Shipley et al. \\
ScH$_6$        & 100 & 144 & -2.0  & Shipley et al. \\
YH$_9$         & 200 & 143 & -3.9  & Shipley et al. \\
MgH$_{13}$     & 300 & 143 & -6.7  & Shipley et al. \\
H$_3$S         & 300 & 143 & -8.1  & JVASP-79487    \\
MgH$_{14}$     & 500 & 143 & -10.8 & Shipley et al. \\
MgH$_{12}$     & 200 & 142 & -3.6  & Shipley et al. \\
SrH$_{15}$     & 500 & 140 & -12.1 & Shipley et al. \\
H$_3$S         & 500 & 138 & -12.6 & JVASP-79487    \\
Na$_2$H$_{11}$ & 100 & 137 & -1.9  & Shipley et al. \\
ScH$_6$        & 200 & 137 & -3.9  & Shipley et al. \\
KH$_{10}$      & 200 & 137 & -6.2  & Shipley et al. \\
CaH$_{15}$     & 200 & 136 & -4.7  & Shipley et al.
\end{tabular}
\caption{A summary of DFT results, including structure (chemical formula), applied pressure, superconducting critical temperature ($T_c$), formation energy per atom (E$_{\textrm{form}}$) at each respective pressure, and the source of the original strucutre (JARVIS-DFT, 3DSC \cite{sommer20223dsc}, or Shipley et al. \cite{PhysRevB.104.054501} dataset) for dynamically stable structures with the highest $T_c$. A continuation of this data is in Table S1. }
\label{table:1}
\end{adjustbox}
\end{table}

\begin{figure}[h]
    \centering
    \includegraphics[trim={0. 0cm 0 0cm},clip,width=0.7\textwidth]{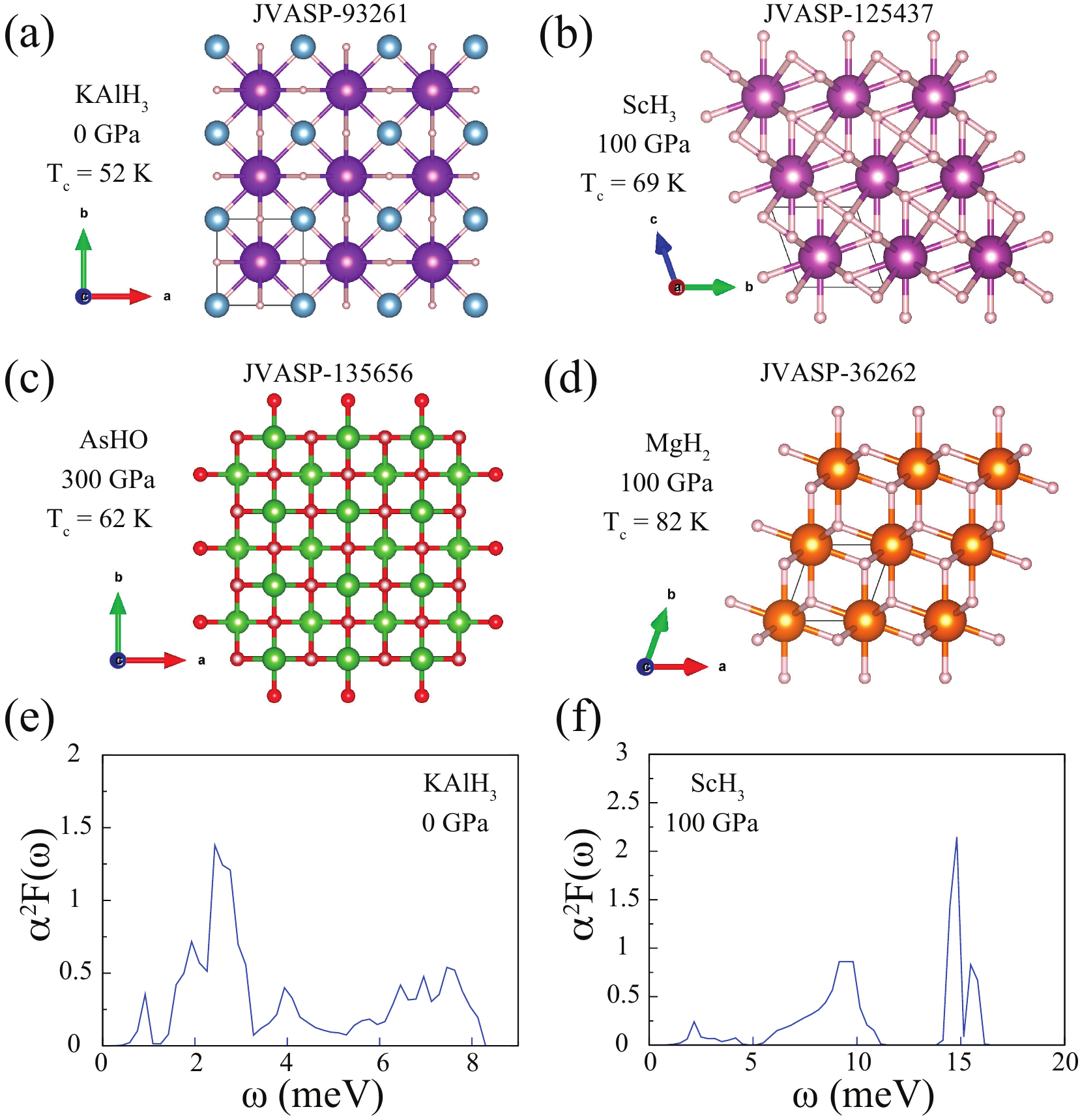}
    \caption{a) - d) Selected structures from JARVIS that were revealed to be superconducting. e) and f) depict selected Eliashberg spectral functions.}
    \label{hydride-jarvis}
\end{figure}

\begin{figure*}[h]
    \centering
    \includegraphics[trim={0. 0cm 0 0cm},clip,width=1.0\textwidth]{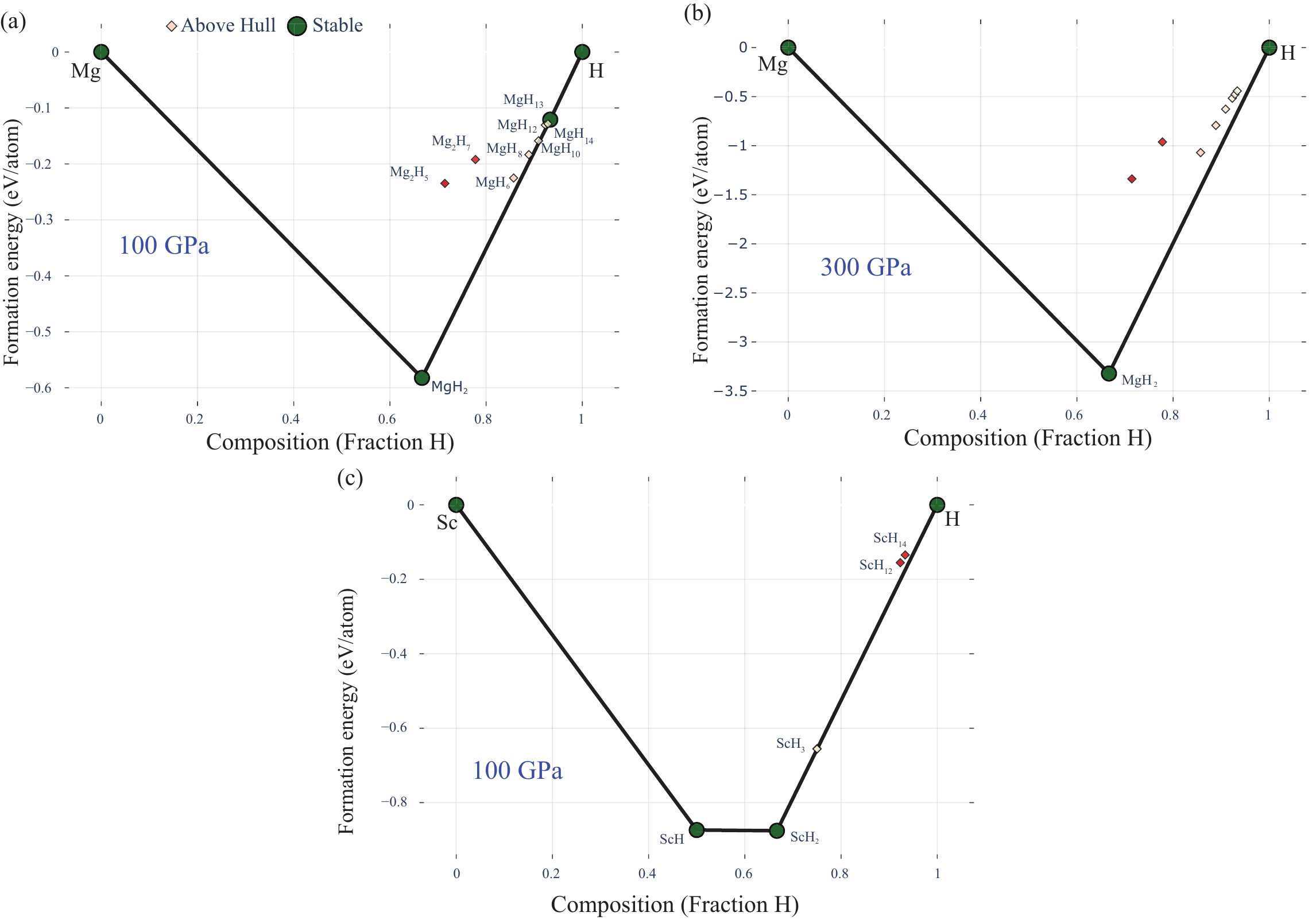}
    \caption{Phase diagrams for some of the selected compounds reported in this work: Mg-H at a) 100 GPa, b) 300 GPa and c) Sc-H at 100 GPa. Formation energy data and code to produce these phase diagrams are available with this manuscript for users to construct their own phase diagrams and add their own DFT data.    }
    \label{hydride-phase}
\end{figure*}

It is clear from Table \ref{table:1} and Table S1 that the majority of the structures that have a high $T_c$ come from the dataset of Shipley et al. \cite{PhysRevB.104.054501}, which were recalculated with our workflow. The main motivation for adding the previously discovered structures from Shipley et al. \cite{PhysRevB.104.054501} was to provide high-$T_c$ data for our deep learning model. All of the structures shown in Table \ref{table:1} and Table S1 are dynamically stable (possess positive phonon frequencies in the phonon density of states). In addition to dynamical stability of these hydride compounds, we assessed the thermodynamic stability of these structures by computing the formation energy of each material (added to Table \ref{table:1} and Table S1). Each formation energy value is computed by subtracting the respective energies of the most energetically favorable elemental solids from the total energy of each hydride compound. For an accurate reference point, we calculated the total energy of each elemental solid under pressure (i.e., to calculate the formation energy of CaH$_6$ at 100 GPa, we subtracted off the energies of elemental Ca and H at 100 GPa). We observe that all of structures in Table \ref{table:1} and most structures in Table S1 possess negative formation energy. By obtaining these formation energy results, we can easily compute the convex hull \cite{ONG2010427,phase} for selected compounds to observe how thermodynamically stable a given phase is with respect to other reported phases (see Fig \ref{hydride-phase}, discussed in the next paragraph). Due to a lack of formation energy data for hydride-based structures at pressures above 0 GPa in most DFT databases \cite{doi:10.1063/1.4812323,oqmd,oqmd-2,choudhary2020joint,10.1063/5.0159299}, constructing a convex hull for a given hydride superconductor has been a challenging task. We hope that by providing a large set of formation energy data for a wide variety of hydride structures under various amounts of pressure can allow others to create their own phase diagrams for compounds of interest (similar to those reported in Fig. \ref{hydride-phase}). We acknowledge that it is possible for a material to exhibit strong electron-phonon coupling and also have a band gap, which would destroy the chances of the candidate material to be superconducting. To investigate this, we calculated the band gap of all of our materials using the PBEsol functional. We found that out of all of the nearly 900 computations we performed, only 41 structures possessed a band gap. Out of these 41 structures, 30 have a band gap and a finite $T_c$ (almost all of these structures have a $T_c$ less than MgB$_2$). These structures are listed in Table S2.

In addition to adding previously discovered high-$T_c$ to our dataset, our workflow revealed some hydride-based structures in JARVIS to be superconducting. These included MgH$_2$ (82 K at 100 GPa), AsHO (62 to 78 K at 200 to 300 GPa), ScH$_3$ (69 K at 100 GPa), and KAlH$_3$ (52 K at 0 GPa). Fig. \ref{hydride-jarvis} depicts some of the selected materials from JARVIS that were revealed to be superconductors, in addition to some of the corresponding Eliashberg spectral functions. Along with this manuscript, all relaxed structures and full DFT calculations are available for further analysis and future utilization. Fig. \ref{hydride-phase} depicts the computed phase diagrams for a few of the selected materials previously mentioned (using pymatgen \cite{ONG2010427,phase}), and we see that all of these structures lie on the respective convex hulls. Specifically in Fig. \ref{hydride-phase} a) - b), we demonstrate how the phase diagram of Mg-H changes with pressure. Although Fig. \ref{hydride-phase} depicts phase diagrams of binary systems, it is possible to construct these phase diagrams for ternary systems using pymatgen. Interestingly, when pressure is added to KAlH$_3$, a metal-to-semiconductor transition is induced from (100 - 500) GPa, quenching the superconducting properties (see Table S2). To our knowledge, there have been no experimental reports of superconductivity in KAlH$_3$. In addition, we found that the SiH$_4$ (identified from 3DSC \cite{sommer20223dsc}) structure possesses a $T_c$ of 71 and 72 for 200 GPa and 300 GPa respectively. It is important to note a different phase of ScH$_3$ has experimentally and theoretically been reported to be superconducting \cite{sch3-exp,doi:10.1073/pnas.0914462107} (with a $Fm\bar{3}m$ space group), but to our knowledge, the phase of ScH$_3$ (with a $Cm$ space group) reported in our manuscript is identified as a superconductor with DFT for the first time. With regards to MgH$_2$, the phase reported in our work has a space group of $Fm-3m$, which differs from the more stable reported $\alpha$-MgH$_2$ with a space group of Pnma \cite{PhysRevB.87.054107}. $\alpha$-MgH$_2$ (Pmna) undergoes a semiconductor to metal transition with higher pressures above 170 GPa, reaching a $T_c$ up to 23 K \cite{PhysRevB.87.054107}. Our reported MgH$_2$ structure ($Fm-3m$), which has been found to be metastable \cite{C8RA07068G,Boonchot_2021}, retains its metallicity from (0 - 500) GPa. To our knowledge, previous studies have not explored the superconducting capabilities of the metastable MgH$_2$ structure (Fm-3m) in detail (besides our previous work at 0 GPa \cite{bulksc}). SiH$_4$ (Silane) is a material that has been previously studied in great detail experimentally and through first-principles, with a variety of phases and pressure-induced phase transitions being reported \cite{PhysRevLett.97.045504,PhysRevB.83.144102,PhysRevLett.101.077002,silane-1,doi:10.1073/pnas.0804148105,PhysRevLett.103.065701,PhysRevB.76.064123}. In this work, we obtained a structure for SiH$_4$ (space group $P2_1/c$) from the 3DSC \cite{sommer20223dsc} database (which maps entries of the experimental Supercon database to Materials Project entries). This structure is found in both JARVIS and Materials Project (JVASP-5281, mp-23739). Although this $P2_1/c$ structure of Silane is insulating at zero pressure, our calculations indicate that the application of large pressure results in an insulator-to-metal transition, making superconductivity possible. Although it has been demonstrated theoretically that the thermodynamic stability of $P2_1/c$ can vary with pressure (i.e., the $C2/c$ phase is more stable than the $P2_1/c$ phase for pressure ranges below 400 GPa \cite{PhysRevLett.97.045504,PhysRevB.83.144102,PhysRevLett.101.077002,silane-1}), we find that $P2_1/c$ Silane possesses strong superconducting properties at (200 - 300) GPa. To our knowledge, DFT calculations for $T_c$ at 200 - 300 GPa have not been carried out for $P2_1/c$ Silane prior to this study and the superconducting properties of $P2_1/c$ Silane have not been experimentally measured at these pressure values. Although our DFT results show dozens of potential high pressure hydride superconductor candidates, very few of our predicted structures have been experimentally realized. In fact, only a handful of materials from our dataset have been synthesized under pressure including H$_3$S \cite{h3s}, LaH$_{10}$ \cite{lah10,https://doi.org/10.1002/anie.201709970,PhysRevLett.122.027001}, CaH$_6$ \cite{cah6}, YH$_9$ \cite{yh9}, and polyhydra of Li-H (LiH$_2$ and LiH$_6$) \cite{PhysRevB.86.064108,doi:10.1073/pnas.1507508112}. In addition, polyhydra of Sr-H \cite{SMITH2009830} and Na-H \cite{nah-poly} with different stoichiometry than our DFT calculations have been synthesized (SrH$_2$, NaH$_3$, NaH$_7$).

\begin{figure}[h]
    \centering
    \includegraphics[trim={0. 0cm 0 0cm},clip,width=1.0\textwidth]{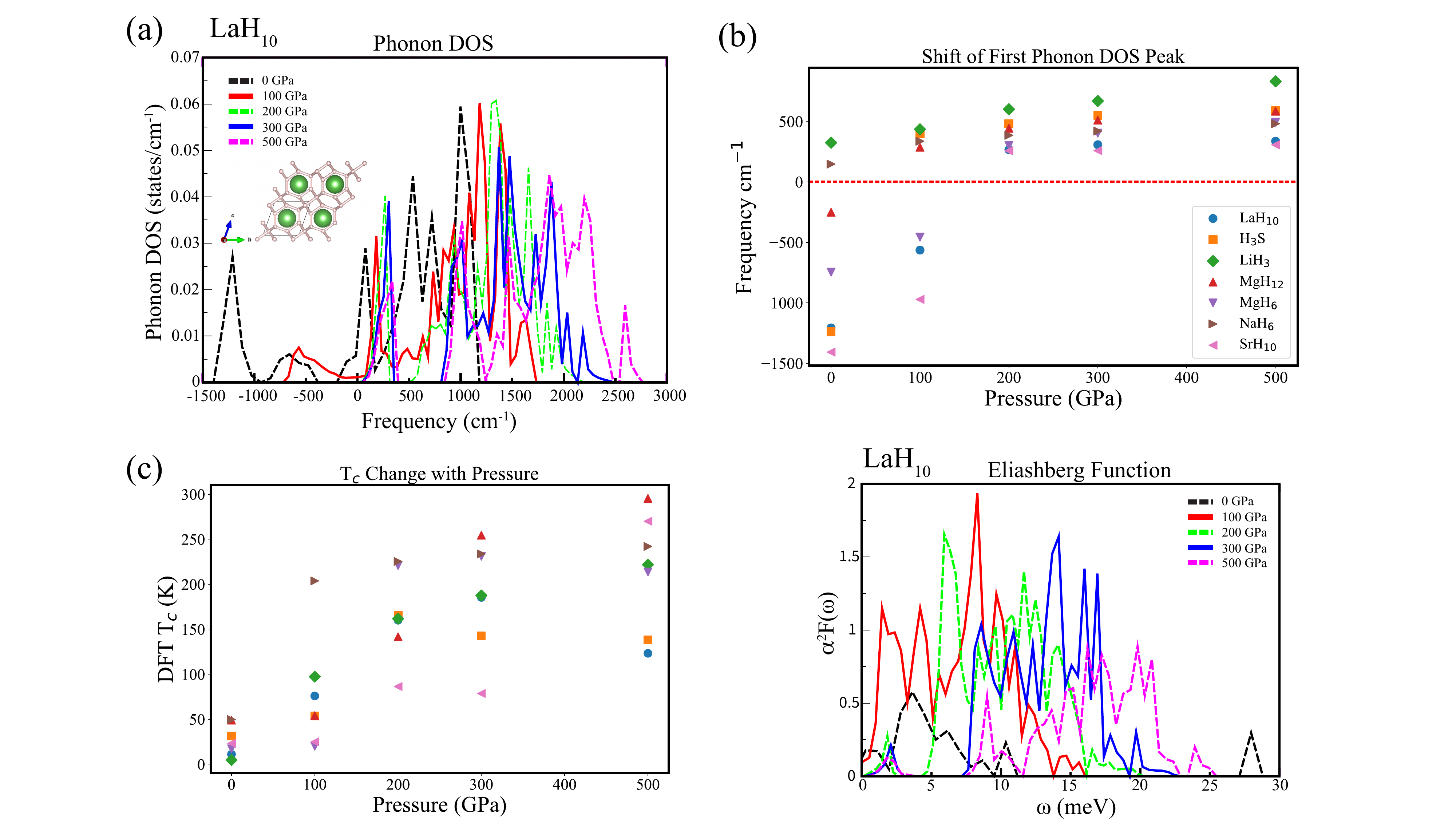}
    \caption{An analysis of DFT results for select materials at various pressures: a) the overlapping phonon density of states for LaH$_{10}$, b) the shift of the first peak in the phonon density of states with respect to applied pressure (points below the red dotted line indicate dynamical instability), c) $T_c$ as a function of pressure, and d) the overlapping Eliashberg function for LaH$_{10}$.  }
    \label{hydride-analysis}
\end{figure}

Our plethora of DFT data on hydride materials under extreme pressures allows us to analyze certain trends in the results. Fig. \ref{hydride-analysis} depicts a thorough analysis of DFT results for selected materials at various pressures. Fig. \ref{hydride-analysis}a)-b) depicts how the phonon density of states changes when pressure is applied. From Fig. \ref{hydride-analysis}a)-b), we observe that the phonon DOS is blueshifted, and the application of high pressure can cause the first peak in the phonon DOS to change from negative to positive (triggering an unstable-to-stable transition), as is the case for LaH$_{10}$, H$_3$S, MgH$_{12}$, MgH$_6$, SrH$_{10}$. Fig. \ref{hydride-analysis}c) depicts how $T_c$ changes with respect to applied pressure. It is clear from the figure that there is not necessarily a linear relationship between applied pressure and an increase in $T_c$. In fact, for certain materials such as LaH$_{10}$ and H$_3$S, $T_c$ increases up to some critical value of pressure and then decreases. This enforces the point that the relationship between superconductivity and applied pressure is entirely material dependent and cannot be described by simple models or equations (i.e., linear or polynomial fitting). Fig. \ref{hydride-analysis}d) depicts the Eliashberg function of LaH$_{10}$ under different amounts of pressure. As expected, the area under the curve is directly proportional to the change in $T_c$ at each pressure value (see Eq. \ref{eq:lambda}).

\subsection{Machine Learning Results}

\begin{figure}[h]
    \centering
    \includegraphics[trim={0. 0cm 0 0cm},clip,width=1.0\textwidth]{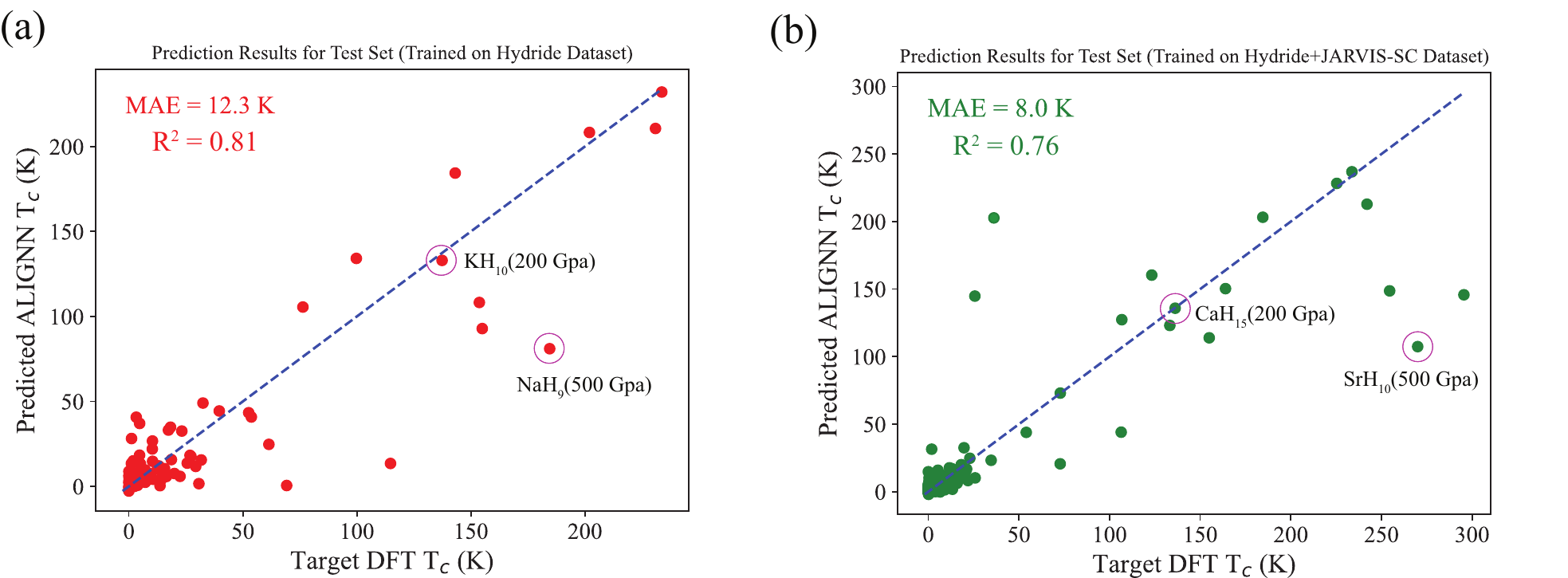}
    \caption{The performance of two separate ALIGNN models for $T_c$ on the test set (random 10 $\%$ of each dataset). a) Is the ALIGNN model trained on the hydride dataset and b) is the ALIGNN model trained on dataset which contains the hydride dataset and the JARVIS superconductor dataset from Ref. \cite{bulksc}. The mean absolute error (MAE) and R$^2$ value are depicted in the inset for each model and the dotted line is shown for reference.}
    \label{alignn-fig}
\end{figure}

The main outputs of this effort are 1) a high quality and diverse set of DFT data for hydride materials under pressure and 2) a well-developed deep learning model to predict $T_c$ at a lower computational cost. The motivation to train a deep learning model to predict $T_c$ for hydrides under various amounts of pressure stems from the fact that 1) each DFT calculation (for the structural relaxation under pressure and the DFPT calculation for the EPC) can be incredibly computationally expensive and 2) (as previously mentioned) there is no clear relationship between superconductivity and applied pressure, where it is entirely material dependent and cannot be described by simple models or equations. To address these two main concerns, we trained an ALIGNN model on our DFT data to predict $T_c$. A similar procedure was carried out in Ref. \cite{bulksc}, where an ALIGNN model was trained on 1000 bulk materials. The $T_c$ model of Ref. \cite{bulksc} achieved an MAE of 1.84 K, but the maximum $T_c$ in this dataset was 33 K. In this work, we trained two separate ALIGNN models on different datasets (the datasets consisted of relaxed structures under pressure and $T_c$ value). One model was trained on purely the hydride dataset (over 900 structures) and the other model was trained on the hydride dataset plus the dataset from Ref. \cite{bulksc} (900+1000 structures). Fig. \ref{alignn-fig}a) and b) depict the performance of both of these ALIGNN models on a random 10 $\%$ test set for each dataset (the random test set is different for each model to minimize bias that arises from partitioning the data into train:test:validation). The MAE for the ALIGNN model trained on the hydride dataset is 12 K and the MAE for the ALIGNN model trained on the hydride plus the JARVIS dataset from Ref. \cite{bulksc} is 8 K. We find the mean absolute deviation (MAD) of the ALIGNN model trained on the hydride dataset to be 32 K while the MAD of the ALIGNN model trained on the hydride plus the JARVIS dataset from Ref. \cite{bulksc} is 19 K. The MAE values of these models are much larger due in part to the fact that the maximum $T_c$ in the datasets are much larger (over 280 K). Fig. \ref{alignn-fig}a) and b) also demonstrates that the addition of data in these deep learning models can reduce the overall MAE. Ideally, if an additional, larger DFT dataset of high-pressure hydride materials was developed (that was consistent with our calculation method, in terms of DFT functional and pseudopotential for consistency), it could be used to augment the training set and could potentially improve performance. Fig. \ref{alignn-fig} also depict some of the outlying ALIGNN predictions on the test set, which can be attributed to random model error. Fig. S1, S2 and S3 depict additional details regarding both ALIGNN models. Fig. S1 shows the training and validation MAE for $T_c$ (K) as a function of the number of epochs. Fig. S2 shows the performance of both ALIGNN models on the training and validation sets. Due to the fact that the MAE values could be somewhat inflated because a large portion of the data lies under 50 K, we plotted the data separately for $T_c$ above 50 K and below 50 K and computed each MAE separately for both ALIGNN models (see Fig. S3). In our previous work in Ref. \cite{bulksc}, we computed $T_c$ by predicting $\lambda$ and $\omega_{\textrm{log}}$ separately with ALIGNN and plugged both quantities into Eq. \ref{eq:mad}. We found that this only resulted in a 4 $\%$ increase in accuracy. For this reason, we decided to directly predict $T_c$ with ALIGNN. We hope that other researchers can utilize this ALIGNN model specifically fine tuned for high-pressure hydrides to screen the broader materials space for high-$T_c$ hydride candidates prior to verifying with DFT (as shown in the workflow in Fig. \ref{hydride-workflow}). It is apparent from Fig. \ref{alignn-fig} and Fig. S3 that the errors in $T_c$ prediction from ALIGNN are relatively high. Despite this limitation of the model, we believe that it can be useful for the classification of potential high-$T_c$ superconductors. The goal is not to outright replace DFT predictions, but to aid in the screening of potential candidates prior to verification by DFT.

Motivated by this idea, we went on to quantify the classification of superconductors for both ALIGNN models. Our first classification metric is whether or not ALIGNN can predict if a structure is superconducting or not ($T_c$ $>$ 0.1 K). We define a false positive as when ALIGNN predicts a $T_c$ $>$ 0.1 K while DFT predicts a $T_c$ $<$ 0.1 K, and a false negative as when ALIGNN predicts a $T_c$ $<$ 0.1 K while DFT predicts a $T_c$ $>$ 0.1 K. In the few cases where ALIGNN erroneously predicts a negative $T_c$ value with magitude larger than 0.1 K (unphysical result), these entries are counted as both false positive and false negative. For the ALIGNN model trained only on the hydride DFT dataset, the test set contained 95 materials. Out of these 95 structures, there were 8 false positives and 4 false negatives (with 3 entries being counted for both categories as unphysical results). For the ALIGNN model trained on the hydride+JARVIS-SC DFT dataset, the test set contained 199 materials. Out of these 199 structures, there were 31 false positives and 9 false negatives (with 8 entries being counted for both categories as unphysical results). Our second classification metric is whether or not ALIGNN can predict if a structure is a high temperature superconducting or not ($T_c$ $>$ 50 K). We define a false positive as when ALIGNN predicts a $T_c$ $>$ 50 K while DFT predicts a $T_c$ $<$ 50 K, and a false negative as when ALIGNN predicts a $T_c$ $<$ 50 K while DFT predicts a $T_c$ $>$ 50 K. For the ALIGNN model trained only on the hydride DFT dataset, we identified 0 false positives and 5 false negatives (16 materials in the test set have a DFT $T_c$ $>$ 50 K). For the ALIGNN model trained on the hydride DFT+JARVIS-SC dataset, we identified 2 false positives and 3 false negatives (17 materials in the test set have a DFT $T_c$ $>$ 50 K). These metrics demonstrate that although the prediction errors are relatively high in Fig. \ref{alignn-fig}, these ALIGNN models can be useful in classification tasks for $T_c$, especially for high-$T_c$ materials. By adding the ALIGNN model to the screening workflow for high pressure hydrides, the number of DFT calculations for superconducting properties can be significantly reduced.

\begin{figure}[h]
    \centering
    \includegraphics[trim={0. 0cm 0 0cm},clip,width=1.0\textwidth]{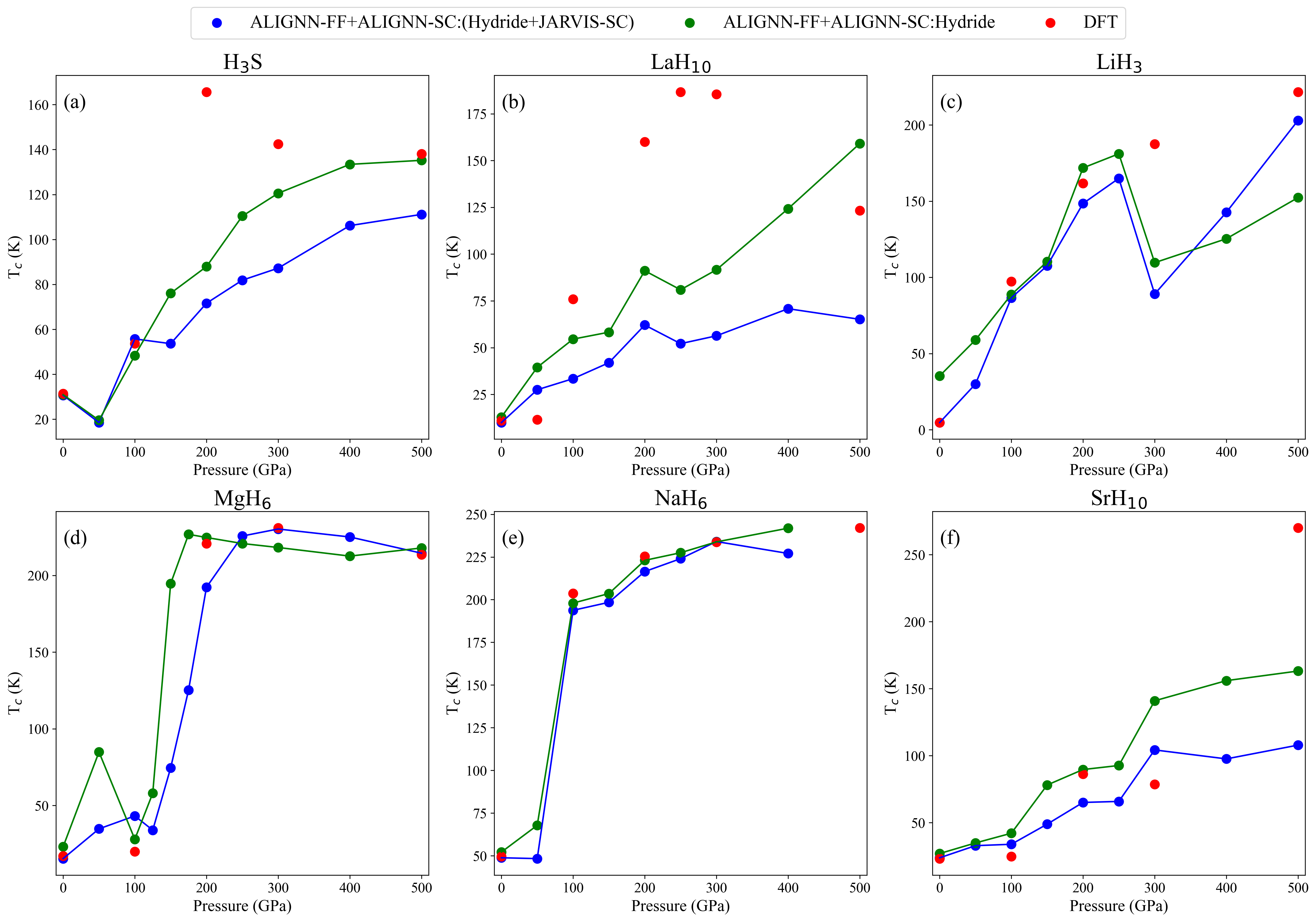}
    \caption{$T_c$ results from DFT (red) along with $T_c$ results from ALIGNN-SC (green for the ALIGNN-SC model from Fig. \ref{alignn-fig}a), blue for the ALIGNN-SC model from Fig. \ref{alignn-fig}b)) that were obtained using ALIGNN-FF relaxed structures at various pressures.   }
    \label{alignn-ff}
\end{figure}

In the previous paragraph, we discussed how ALIGNN trained on DFPT calculations for $T_c$ can accelerate the prediction of $T_c$. Although this is an extremely useful tool, the ALIGNN model for $T_c$ assumes that the relaxed structure under pressure is provided. In addition, if only the relaxed structure with zero applied pressure is provided, it will usually yield a low value for $T_c$, which may result in the structure being disregarded as a candidate for a potential superconductor. Although the DFPT computation of the EPC is much more computationally expensive than the DFT structural relaxation under pressure, this relaxation can still be a time consuming endeavor. One way to circumvent this and eliminate the cost of DFT structural relaxation is to use a universal machine learned force field (such as JARVIS-Force-Field \cite{choudhary2023unified}, M3GNET/MatGL \cite{m3net} and GemNet-OC \cite{gasteiger2022gemnetoc}) to relax the structure prior to ALIGNN prediction of $T_c$. In this work, we used the pretrained ALIGNN-Force-Field (FF) \cite{choudhary2023unified} (universal machine learning force field) with the Atomic Simulation Environment (ASE) \cite{Hjorth_Larsen_2017} FIRE \cite{PhysRevLett.97.170201} optimizer to relax the hydride-based structures under various amounts of pressure. Pressure was induced in each structure by compressing the volume and allowing the atoms and bond angles to relax under that constant volume value. ALIGNN-FF was developed to handle chemically and structurally diverse crystalline systems, where the entirety of the previous JARVIS-DFT dataset was used as training, (which contains 4 million energy-force entries for 89 elements of the periodic table, 307113 which are used for training) \cite{choudhary2023unified}. It is important to note that none of the high pressure calculations in this work were used to train this ALIGNN-FF.

The ALIGNN-FF+ALIGNN-Superconducting (ALIGNN-SC) results along with DFT are depicted in Fig. \ref{alignn-ff}. Specifically, we tested both ALIGNN-SC models (one model trained on the hydride dataset, the other trained on the hydride dataset plus the dataset from Ref. \cite{bulksc}) on the relaxed structures from ALIGNN-FF under pressure. As seen in Fig. \ref{alignn-ff} there is good qualitative agreement between DFT and ALIGNN and meaningful trends can be extracted from the purely machine learning results. Regardless if the structures under pressure were in the training, validation or test sets during ALIGNN training, there were no DFT calculations performed at intermediate pressure values such as (50, 150, 250, and 400) GPa (with the exception of DFT calculations for LaH$_{10}$ at 50 GPa and 250 GPa). From Fig. \ref{alignn-ff}, we observe that ALIGNN-FF+ALIGNN-SC can interpolate reasonably well to intermediate values of pressure that were not part of the original dataset, which allows us to reliably compute $T_c$ for a larger number of pressure values and obtain a finer map of how superconducting properties change with pressure, avoiding unnecessary DFT calculations. This can also give us an indication of where the critical pressure value lies (pressure value which results in the maximum value of $T_c$) and is a cheaper alternative to brute-force DFT calculations at several values of pressure.

\section{Conclusion}
In this work, we have computed the superconducting properties of over 900 hydride-based materials with DFT under ultrahigh pressures (0 GPa to 500 GPa). In addition to adding previously discovered high-$T_c$ materials to our dataset and revealing previously undiscovered superconductors from the JARVIS database, we trained an ALIGNN model to predict $T_c$ with reasonable accuracy. We took this one step further and coupled our ALIGNN model for $T_c$ to the ALIGNN-FF universal force-field to relax the structures with minimal computational cost prior to ALIGNN $T_c$ prediction. By utilizing these deep learning tools for structural relaxation under pressure and property prediction, enhanced materials screening can be enabled and the number of DFT calculations (and overall computational cost) can be significantly reduced, allowing for a broader search of material space for new high pressure hydride superconductors.  

\section{Future Perspectives}
Data-driven discovery and design of novel high pressure hydride superconductors is a rapidly changing and exciting field of physics and materials science. Coupling high-throughput first-principles (DFT) simulations with machine learning techniques can allow for a more broad search of the high pressure superconductor landscape. Most importantly, these machine learning models can be used as effective classification tools to screen candidate systems prior to more in-depth theoretical calculations and eventual experimental synthesis. In order for this field to progress, the amount of publicly available and unique high-throughput DFT data must increase, which can be one of the most effective routes to improving the accuracy of these machine learning models. Collaborative efforts such as the JARVIS-Leaderboard \cite{leaderboard}, which allows users to view and contribute materials science data and machine learning models and assess accuracy/fidelity (while enhancing transparency and reproducibility) and OPTIMADE (Open Databases Integration for Materials Design) \cite{D4DD00039K}, which allows for the easy and interoperable access to information across different materials databases in a uniform format can play an important role in granting access to relevant data and metrics with the goal of advancing the field of materials design.

 \section{Data Availability Statement}
Software packages mentioned in the article can be found at \url{https://github.com/usnistgov/jarvis}. All code and data specific to this work can be found at \url{https://doi.org/10.6084/m9.figshare.25270012} and model benchmarks can be found at \url{https://pages.nist.gov/jarvis_leaderboard/}.    

 \section{Notes}
Please note that the use of commercial software (VASP) does not imply recommendation by the National Institute of Standards and Technology.

 \section{Conflicts of Interest}
The authors declare no competing interests. 
 
\section{Acknowledgments}
All authors thank the National Institute of Standards and Technology for funding, computational, and data-management resources.

\bibliography{Main}

\end{document}


\maketitle

\begin{table}[]
\begin{adjustbox}{width=0.5\textwidth}
\small
\begin{tabular}{l|l|l|l|l}
Structure    & Pressure (GPa) & $T_c$ (K) & E$_{\textrm{form}}$ & Source of Structures \\
\hline
Na$_2$H$_{11}$ & 200 & 136 & -5.5  & Shipley et al.              \\
Na$_2$H$_{11}$ & 200 & 133 & -5.5  & Shipley et al.              \\
Na$_2$H$_{11}$ & 300 & 133 & -8.1  & Shipley et al.              \\
Na$_2$H$_{11}$ & 100 & 132 & -1.9  & Shipley et al.              \\
ScH$_6$        & 300 & 132 & -5.7  & Shipley et al.              \\
ScH$_{12}$     & 300 & 129 & -5.8  & Shipley et al.              \\
KH$_{10}$      & 100 & 127 & -2.9  & Shipley et al.              \\
Na$_2$H$_{11}$ & 300 & 127 & -8.1  & Shipley et al.              \\
ScH$_6$        & 500 & 127 & -12.4 & Shipley et al.              \\
MgH$_{10}$     & 200 & 124 & -3.6  & Shipley et al.              \\
LaH$_{10}$     & 500 & 123 & -9.4  & JVASP-149370                \\
ScH$_{12}$     & 200 & 119 & -3.9  & Shipley et al.              \\
CaH$_6$        & 500 & 119 & -13.1 & Shipley et al.              \\
H              & 50  & 117 & 0.2   & JVASP-21216                 \\
H              & 50  & 115 & 0.2   & JVASP-25330                 \\
Na$_2$H$_{11}$ & 0   & 113 & 0.3   & Shipley et al.              \\
MgH$_{14}$     & 100 & 113 & -1.8  & Shipley et al.              \\
MgH$_{8}$      & 500 & 108 & -11.4 & Shipley et al.              \\
SrH$_{15}$     & 300 & 107 & -5.7  & Shipley et al.              \\
ScH$_{12}$     & 100 & 102 & -2.0  & Shipley et al.              \\
MgH$_8$        & 300 & 100 & -7.1  & Shipley et al.              \\
LiH$_3$        & 100 & 97  & -1.8  & Shipley et al.              \\
NaH$_5$        & 500 & 94  & -12.3 & Shipley et al.              \\
NaH$_5$        & 200 & 87  & -5.7  & Shipley et al.              \\
CaH$_{10}$     & 200 & 84  & -5.1  & Shipley et al.              \\
MgH$_8$        & 200 & 84  & -3.5  & Shipley et al.              \\
MgH$_2$        & 100 & 82  & -1.7  & JVASP-36262                 \\
SrH$_{10}$     & 300 & 79  & -5.7  & Shipley et al.              \\
AsHO           & 300 & 78  & -2.4  & JVASP-135656                \\
NaH$_5$        & 300 & 77  & -8.3  & Shipley et al.              \\
ScH$_{14}$     & 300 & 76  & -5.8  & Shipley et al.              \\
SrH$_{15}$     & 200 & 74  & -3.9  & Shipley et al.              \\
Mg$_2$H$_7$    & 100 & 73  & -1.7  & Shipley et al.              \\
SiH$_4$        & 200 & 72  & -3.2  & 3DSC (JVASP-5281, mp-23739) \\
LiH$_6$        & 100 & 71  & -1.9  & Shipley et al.              \\
SiH$_4$        & 300 & 71  & -4.9  & 3DSC (JVASP-5281, mp-23739) \\
Mg$_2$H$_7$    & 500 & 71  & -13.0 & Shipley et al.              \\
ScH$_3$        & 100 & 69  & -2.6  & JVASP-125437                \\
NaH$_5$        & 100 & 67  & -1.9  & Shipley et al.              \\
Mg$_2$H$_5$    & 100 & 67  & -1.6  & Shipley et al.              \\
CaH$_{15}$     & 100 & 66  & -2.6  & Shipley et al.              \\
LaH$_7$        & 200 & 66  & -3.8  & Shipley et al.              \\
AsHO           & 200 & 62  & -1.6  & JVASP-135656                \\
MgH$_{14}$     & 0   & 61  & 0.2   & Shipley et al.              \\
MgH$_{14}$     & 200 & 60  & -3.6  & Shipley et al.              \\
Mg$_2$H$_7$    & 200 & 55  & -3.4  & Shipley et al.              \\
MgH$_{12}$     & 100 & 54  & -1.7  & Shipley et al.              \\
H$_3$S         & 100 & 54  & -1.5  & JVASP-79487                 \\
NaH$_6$        & 0   & 53  & 0.2   & Shipley et al.              \\
KAlH$_3$       & 0   & 52  & -0.1  & JVASP-93261                 \\
NaH$_6$        & 0   & 50  & 0.2   & Shipley et al.              \\
ScH$_{14}$     & 100 & 48  & -2.0  & Shipley et al.              \\
Mg$_2$H$_7$    & 300 & 46  & -8.7  & Shipley et al.              \\
MgH$_8$        & 100 & 43  & -1.7  & Shipley et al.              \\
Mg$_2$H$_5$    & 200 & 43  & -3.2  & Shipley et al.              \\
SrH$_{15}$     & 100 & 41  & -2.1  & Shipley et al.              \\
LiH$_2$        & 100 & 40  & -1.7  & Shipley et al.             
\end{tabular}
\caption{A continuation of DFT results from Table 1, including structure (chemical formula), applied pressure, superconducting critical temperature ($T_c$), formation energy per atom (E$_{\textrm{form}}$) at each respective pressure, and the source of the original strucutre (JARVIS-DFT, 3DSC \cite{sommer20223dsc}, or Shipley et al. \cite{PhysRevB.104.054501} dataset) for dynamically stable structures with the highest $T_c$.}
\label{table:2}
\end{adjustbox}
\end{table}

\begin{table}[]
\begin{tabular}{lrrrl}
Structure & Pressure (GPa) & $T_c$ (K) & E$_{\textrm{gap}}$ (eV) & Source of Structures \\
\hline
HIO       & 0                                  & 2                          & 0.2                               & JVASP-118424         \\
BaH$_3$      & 0                                  & 11                         & 1.3                               & JVASP-125376         \\
CaH$_3$      & 0                                  & 8                          & 0.5                               & JVASP-125381         \\
SrH$_3$      & 0                                  & 9                          & 1.2                               & JVASP-125442         \\
NaHS      & 0                                  & 1                          & 0.1                               & JVASP-135011         \\
BaH$_3$      & 0                                  & 10                         & 1.3                               & JVASP-135869         \\
RbMgH$_2$    & 0                                  & 5                          & 2.0                               & JVASP-139042         \\
KH$_{10}$      & 0                                  & 10                         & 4.5                               & Shipley et al.       \\
LiH$_3$      & 0                                  & 5                          & 1.8                               & Shipley et al.       \\
MgH$_{10}$     & 0                                  & 2                          & 1.4                               & Shipley et al.       \\
MgH$_{13}$     & 0                                  & 27                         & 0.4                               & Shipley et al.       \\
NaH$_6$      & 0                                  & 53                         & 0.5                               & Shipley et al.       \\
GeHO$_2$     & 100                                & 19                         & 1.2                               & JVASP-114288         \\
Li$_2$H      & 100                                & 12                         & 0.8                               & JVASP-116261         \\
BH$_2$O$_2$     & 100                                & 7                          & 0.8                               & JVASP-119915         \\
KAlH$_3$     & 100                                & 34                         & 2.2                               & JVASP-93261          \\
GeHO$_2$     & 200                                & 22                         & 1.9                               & JVASP-114288         \\
BH$_2$O$_2$     & 200                                & 1                          & 0.9                               & JVASP-119915         \\
CaB$_2$H$_2$    & 200                                & 28                         & 0.4                               & JVASP-36669          \\
ZrH$_3$      & 200                                & 1                          & 0.3                               & JVASP-39635          \\
KAlH$_3$     & 200                                & 10                         & 3.8                               & JVASP-93261          \\
BH$_2$O$_2$     & 250                                & 2                          & 1.2                               & JVASP-119915         \\
Zr$_2$HN$_2$Cl$_2$ & 300                                & 16                         & 0.1                               & JVASP-104666         \\
GeHO$_2$     & 300                                & 29                         & 2.3                               & JVASP-114288         \\
BH$_2$O$_2$     & 300                                & 4                          & 1.7                               & JVASP-119915         \\
KAlH$_3$     & 300                                & 20                         & 4.9                               & JVASP-93261          \\
GeHO$_2$     & 500                                & 20                         & 2.8                               & JVASP-114288         \\
Li$_2$H      & 500                                & 28                         & 1.1                               & JVASP-116261         \\
BH$_2$O$_2$     & 500                                & 63                         & 4.9                               & JVASP-119915         \\
KAlH$_3$     & 500                                & 19                         & 4.2                               & JVASP-93261    \\     
\end{tabular}
\caption{A list of structures that possess strong electron-phonon coupling and a band gap, including structure (chemical formula), applied pressure, superconducting critical temperature ($T_c$), and the source of the original strucutre (JARVIS-DFT or Shipley et al. \cite{PhysRevB.104.054501} dataset).}
\label{table:gap}
\end{table}

\begin{figure}[h]
    \centering
    \includegraphics[trim={0. 0cm 0 0cm},clip,width=1.0\textwidth]{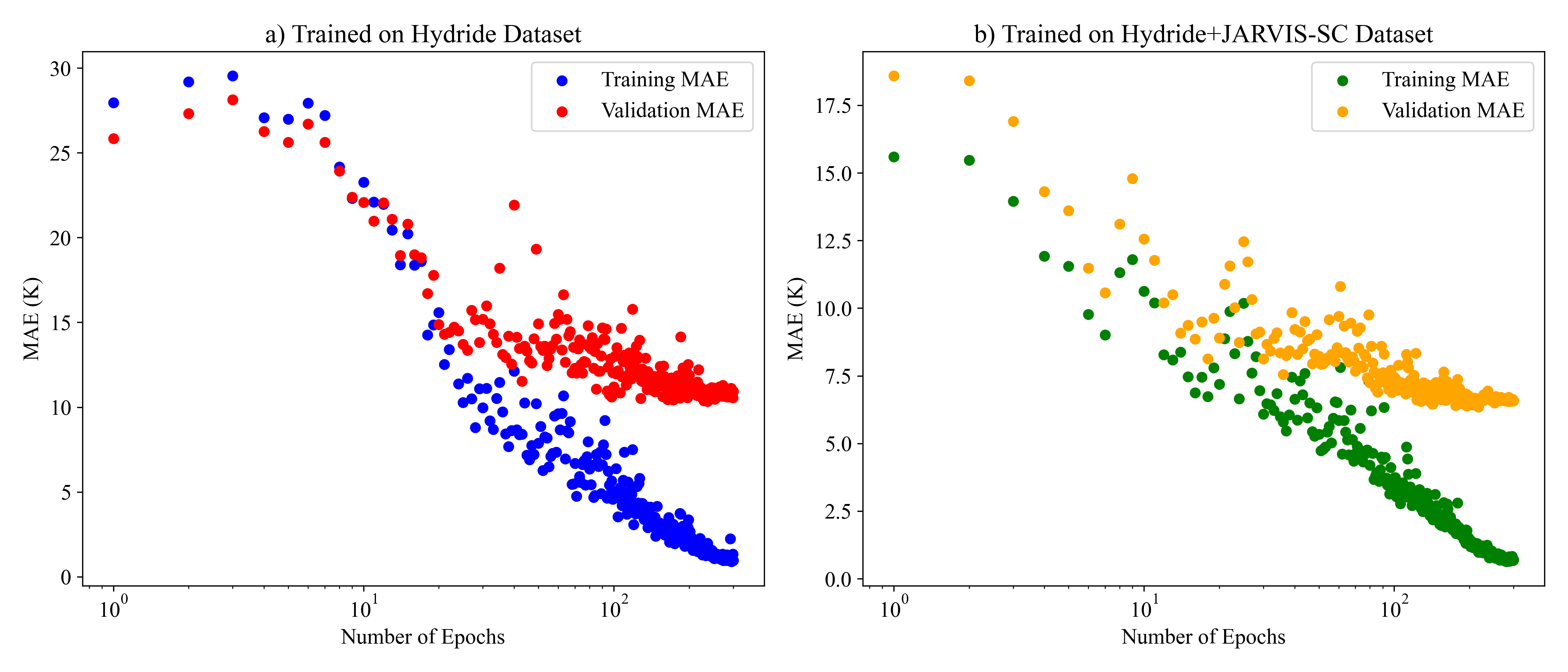}
    \caption{The training and validation MAE for $T_c$ (K) as a function of the number of epochs for both ALIGNN models displayed in Fig. 6. }
    \label{mae-train}
\end{figure}

\newpage
\begin{figure}[h]
    \centering
    \includegraphics[trim={0. 0cm 0 0cm},clip,width=1.0\textwidth]{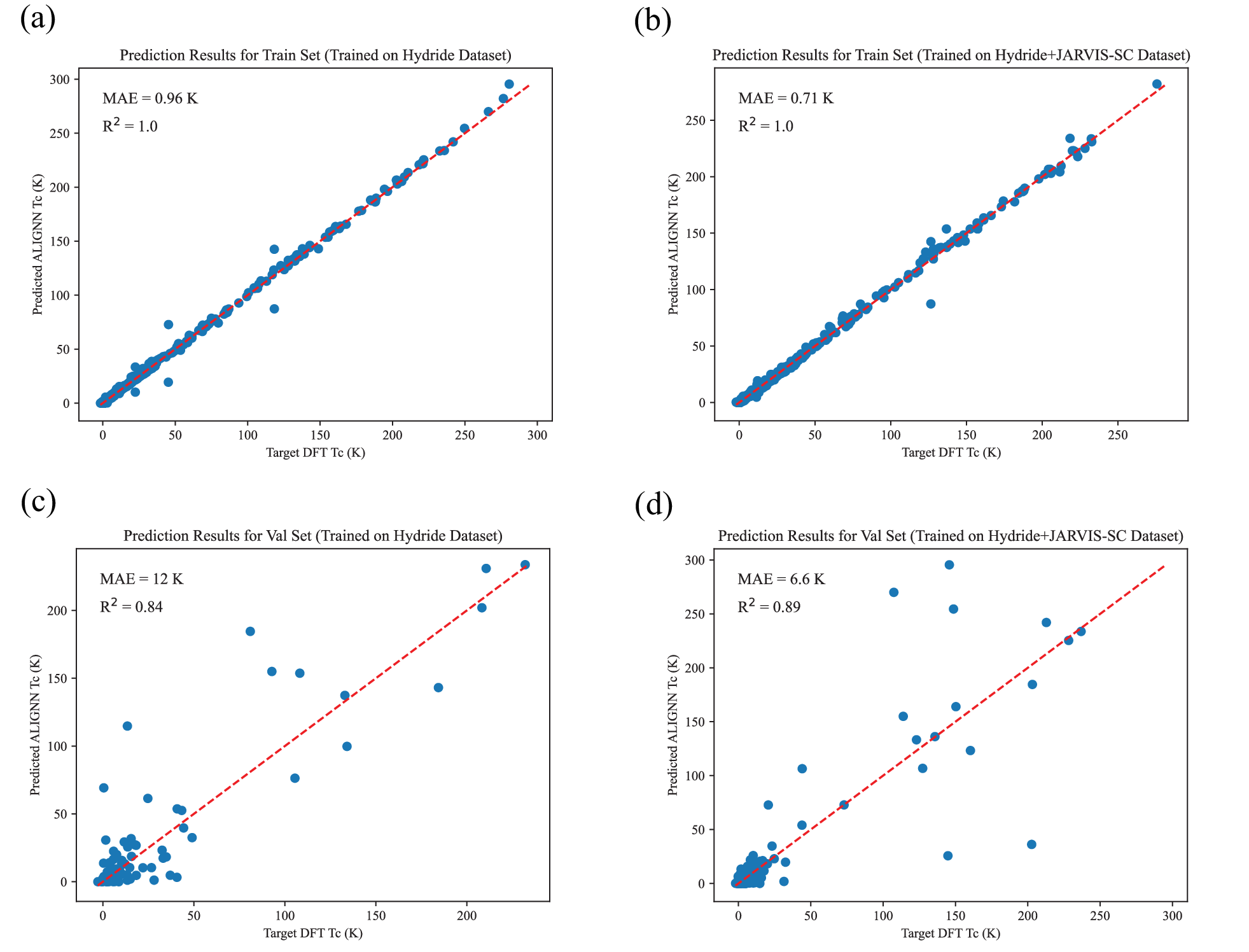}
    \caption{The performance of two separate ALIGNN models for $T_c$ on the training set (random 80 $\%$ of each dataset) and validation set (random 10 $\%$ of each dataset). a) and c) are the ALIGNN models trained on the hydride dataset and b) and d) are the ALIGNN models trained on dataset which contains the hydride dataset and the JARVIS superconductor dataset from Ref. \cite{bulksc}. The mean absolute error (MAE) and R$^2$ value are depicted in the inset for each model and the dotted line is shown for reference.}
    \label{train_val}
\end{figure}

\newpage
\begin{figure}[h]
    \centering
    \includegraphics[trim={0. 0cm 0 0cm},clip,width=1.0\textwidth]{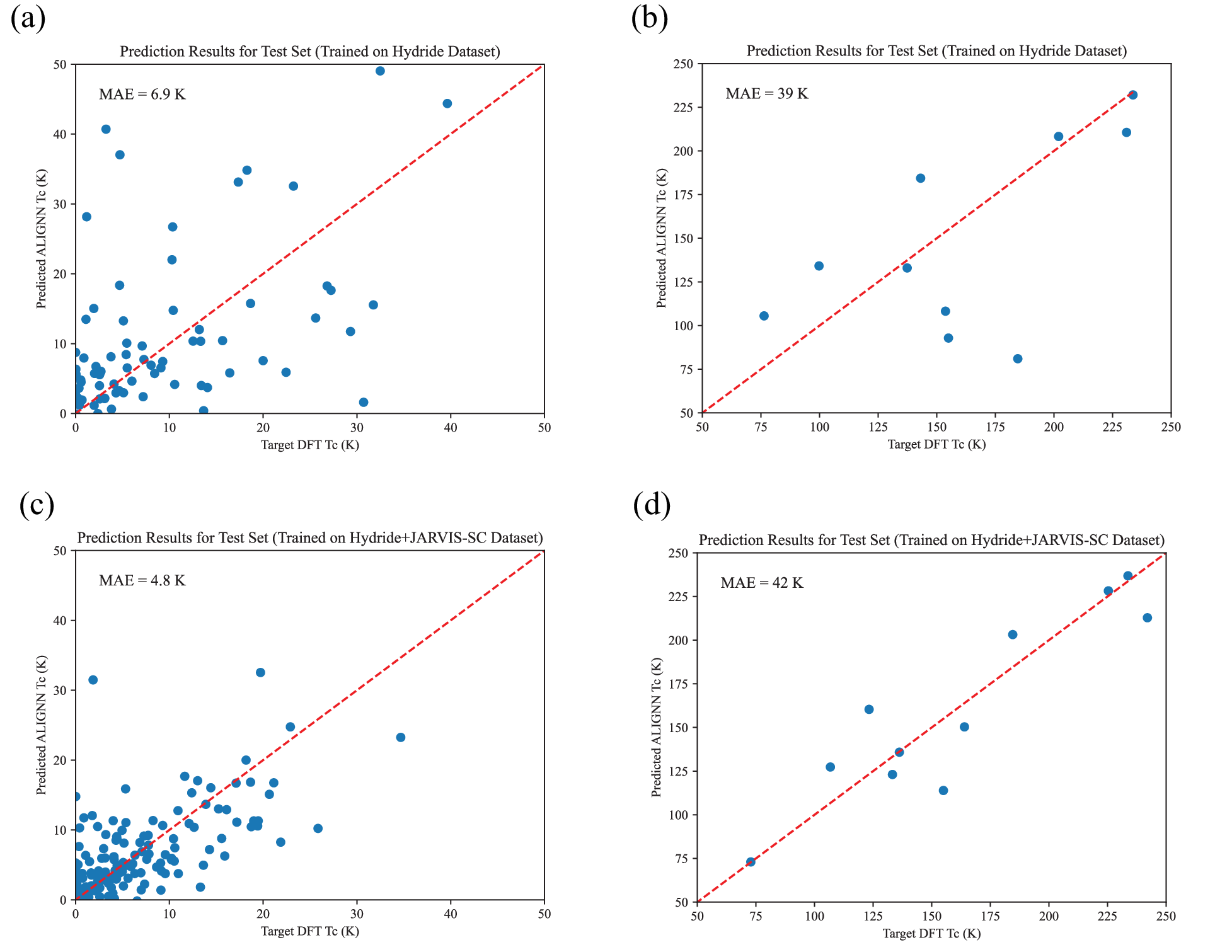}
    \caption{The results of the ALIGNN predictions on the test set (in Fig. 6) plotted in two separate ranges, above and below 50 K. The separate MAE for each range of data is depicted in the inset.  }
    \label{test_range}
\end{figure}

\newpage
\bibliography{Main}